\documentstyle[11pt]{article}
\begin{document}
\begin{titlepage}
\begin{flushright}
{\large \bf PKU-TH-96-2 \\
  Ms.Code:DN5705}
\end{flushright}
\vspace{0.3in}
\begin{center}
{\LARGE Yukawa corrections to top pair production in\\
photon-photon collision}
\vspace{.5in}

  Chong Sheng Li$^{a,b,e}$,
  Jin Min Yang$^{a,c}$, Yun-lun Zhu $^{a,b,f}$ and Hong Yi Zhou$^{a,d}$ \\
\vspace{.5in}
$^a$ CCAST(World Laboratory)\\ 
P.O.Box 8730, Beijing 100080, P.R.China\\
$^b$ Department of Physics, Peking University,\\
Beijing 100871, P.R.China$^*$\\
$^c$ Physics Department, Henan Normal University,\\
Xin Xiang, Henan 453002, P.R.China$^*$\\
$^d$ Physics Department, Tsinghua University,\\
Beijing 100084, P.R.China$^*$\\
$\ \ \ \ \ \ ^e$ E-mail: csli@svr.bimp.pku.edu.cn\\
 $^f$ E-mail: zhuyl@sun.ihep.ac.cn\\
 
\vspace{.3in}
\end{center}
\begin{footnotesize}
\begin{center}\begin{minipage}{5in}

\begin{center} ABSTRACT\end{center}

~~~The $O(\alpha m_t^2/m_W^2)$ Yukawa corrections to top pair production
in photon-photon collision are calculated in the standard model (SM),
the general two-Higgs-doublet model (2HDM) as well as the minimal 
supersymmetric model (MSSM). We found that the correction to the
cross section can only reach a few percent in the SM, but can be quite
significant ($>$10\%) in the 2HDM and MSSM for favorable parameter values,
which may be observable at the high energy $e^+e^-$ colliders.

\end{minipage}\end{center}
\end{footnotesize}
\vspace{.4in}

~~PACS number: 14.65.Ha; 12.38Bx; 14.80.Cp

\vspace{0.2in}

$^*\ \ $ mailing address. 
\vfill
\end{titlepage}
\eject
\rm

\begin{center}1. Introduction \end{center}

   Recently, the evidence for top quark production, with a mass of 
$176\pm 8(stat)\pm 10(syst) $ GeV and $199^{+19}_{-21}(stat)\pm 22 (syst)$
GeV  has been reported by the CDF and D0 collaboration, respectively [1].
Due to its large mass, the discovery of the top quark will
open a number of new and interesting issues, such as the precision measurement
of the mass, width and Yukawa couplings of the top quark through its direct production
and subsequent decay at both hadron and $e^+e^-$  colliders. But even with
$1000 pb^{-1}$  of luminosity, the Fermilab Tevatron could determine the top
mass to $ 5$ GeV or better[2]. At the future multi-TeV
proton colliders such as the CERN Large Hadron Collider (LHC), $t\overline t$
production will be enormously larger than the Tevatron rates, but the accuracy
with which the top mass can be measured in proton colliders is limited to about
$2\sim 3$ GeV[2].  Bloude {\it et al.}[3], have argued that one must know the top mass
to $1$ GeV  to take full advantage of the constraints that precision electroweak
measurements put on the Higgs boson and other massive particles which might
contribute to electroweak loops. Beyond this, it would be wonderful to make a
precision measurement fo the basic parameter $m_t$  to $0.3 $GeV  or better for
looking for new physics beyond the SM by the loop processes which are sensitive
to $m_t$.  At the next-generation linear collider (NLC) operating at a
center-of-mass energy of $500$ GeV-$2000$ GeV  with a luminosity of the order of
$10^{33}cm^{-2}s^{-1}$,  the $e^+e^-\rightarrow t\overline t$  events rate
would be around $10^4/yr$,  comparable with the Tevatron, however the events
would be much cleaner and top parameters would be easier to extract. At the
NLC a top mass measurement with statistical uncertainty $0.3$ GeV from $10 fb^{-1}$
luminosity is expected[2] and it is possible to separately measure all of the
various production and decay from factors of the top quark at the level of
a few percent[4].

   Nowadays, the possibility of transforming a linear $e^+e^-$ collider into a
$\gamma\gamma $  collider deserves a lot of attention. With the advent of the
new collider technique[5] the collision of high energy, high intensity photon
beams , obtained by using the old idea of Compton laser backscattering[6],
can be realized in the NLC. The back-scattering of laser photons off the
colliding electron and positron beams can yield intense and energetic photon
beams which then collide with the high luminosity. There are many uses for such
high photon-photon luminosity, one of the most important may be for the production
of top quark paris. It has been found[7] that $t\overline t$ production in 
$\gamma\gamma $  collisions realized by laser back-scattering is slightly larger
than the direct $e^+e^- \rightarrow t\overline t$  production for $m_t<130$ GeV
at $\sqrt{s}=0.5TeV$ , and
at $\sqrt{s}=1TeV$ the production of $\gamma\gamma\rightarrow
t\overline t$ is much larger than $e^+e^- \rightarrow t\overline t$ for $m_t\sim
100-200$ GeV both with and without considering the threshold QCD effect.
In the SM, the cross section for top quark pair productions in $\gamma \gamma$ collisions
have been calculated with higher order QCD correction[8].
The radiative corrections to $\gamma \gamma t\overline t$ from final state
Higgs exchange interactions has also computed in Ref.[9]. The correction is of
order $O(2-4\%)$ for typical values of the Higgs boson mass and top quark mass.
In this paper we calculate the $O(\alpha m_t^2/m_W^2)$ Yukawa correction
in a two Higgs doublet model (2HDM)(Model II)[10] and in the minimal 
supersymmetric model(MSSM) , in which there are three neutral and two charged physical
Higgs bosons, $~H,h,A, H^{\pm}$, of which $H$ and $h$ are CP-even and $A$ is
CP-odd. 
The $O(\alpha m_t^2/m_W^2)$ Yukawa correction arise from the virtual
effects of the third family (top and bottom) quarks, charged and neutral Higgs
bosons, as well as the Goldstone bosons ($G^0, G^{\pm}$).
The results of the standard model can be obtained from our calculations
as a special case.
In Sec. II, we present the analytic results in terms of the well-known
standard notation of one-loop Feynman integrals. In Sec. III, we present
some numerical examples and discuss the implication of our results.
And in the appdendix we list the form factors in the cross section.

\begin{center} 2. Calculations \end{center}

The  relavent Feynman diagrams are shown in Fig.1 and the  Feynman rules
can be found in Ref.[10]. In our calculation, we use dimensional
regularization to regulate all the ultraviolet divergences in the virtual
loop corrections and we adopt the on-mass-shell renormalization scheme[11].
In our calculations we keep the term $m_b \tan\beta$ in the
in the charged Higgs couplings to third family quarks since its effects
become rather important for large $\tan\beta$. 
Taking into account
the $O(\alpha m_t^2/m_W^2)$ Yukawa corrections, the renormalized
amplitude for $\gamma \gamma \rightarrow t \bar t$
is given by
\begin{eqnarray}
 M_{ren}&=&M_{ren}^{(t)}+M_{ren}^{(u)},\\
 M_{ren}^{(t)}&=&M_{0}^{(t)}+\delta { M}^{(t)}\nonumber\\
	      &=&M_{0}^{(t)}+\delta { M}^{s(t)}
	    +\delta { M}^{v(t)}+\delta { M}^{b(t)}
            +\delta {M}^{\Delta(t)},\\
 M_{ren}^{(u)}&=& M_{ren}^{(t)}(p_3 \leftrightarrow p_4,
 \hat{t} \rightarrow \hat{u}),
\end{eqnarray}
where $ M_{0}$ is the amplitude at tree level,
$\delta { M}^{s}, \delta { M}^{v}$, $\delta { M}^{b}$ and $\delta {M}^{\Delta(t)}$
represent the $O(\alpha m_t^2/m_W^2)$ Yukawa corrections arising
from the self energy diagram Fig.1(d), vertex diagram Fig.1(f)-(i) 
, box diagrams Fig.1(l)-(n) and digrams Fig.1(j),(k), respectively.
$\hat{t}=(p_4-p_2)^2$, $\hat{u}=(p_1-p_4)^2 $ and
$p_3(p_4)$ denote the momentum of the two incoming photons, and
$p_2(p_1)$ are momentum of the outgoing t quark and its
antiparticle.
The explicit forms of these matrix elements are given by

\begin{eqnarray}
M_{0}^{(t)}&=&-i \frac{e^2}{\hat{t}-m_t^2}
	\epsilon_{\mu}(p_4)\epsilon_{\nu}(p_3){\bar u}(p_2)~\gamma^{\mu}~
 (\gamma \cdot p_3-\gamma \cdot p_1+m_t)\gamma^{\nu}{ v}(p_1),\\
\delta M^{s(t)}&=&\!ie^2Q_t^2\frac{\alpha m_t^2}{16\pi m_W^2s_W^2}
   \epsilon^{\mu}(p_4)\epsilon^{\nu}(p_3)\bar{u}(p_2)
    (f_2^s\gamma_{\mu}\gamma_{\nu}\!+\!f_6^sp_{2\mu}\gamma_{\nu}\!+\!
     f_{12}^s\rlap/p_4 \gamma_{\mu}\gamma_{\nu})v(p_1)\\
\delta M^{v(t)}&=&ie^2Q_t^2\frac{\alpha m_t^2}{16\pi m_W^2s_W^2}
    \epsilon^{\mu}(p_4)\epsilon^{\nu}(p_3)\bar{u}(p_2)
(f_2^v~\gamma_{\mu}\gamma_{\nu} +f_3^v~p_{1\nu}~\gamma_{\mu}
   +f_6^v~p_{2\mu}~\gamma_{\nu}\nonumber\\
& & +f_{12}^v~\rlap/p_4 \gamma_{\mu}\gamma_{\nu}
    +f_{13}^v~\rlap/p_4 p_{1\nu}\gamma_{\mu}+f_{16}^v~\rlap/p_4
 p_{2\mu}\gamma_{\nu})v(p_1),\\
\delta M^{b(t)}& =& ie^2Q_t^2\frac{\alpha m_t^2}{16\pi m_W^2 s_W^2}
    \epsilon_{\mu}(p_4)\epsilon_{\nu}(p_3)\bar u(p_2)~\left [
    f_1^{b}g^{\mu\nu}+f_2^{b}\gamma^{\mu}\gamma^{\nu} \right.\nonumber\\
& & +f_3^{b}p_1^{\nu}\gamma^{\mu}
    +f_4^{b}p_1^{\mu}\gamma^{\nu}
    +f_5^{b}p_2^{\nu}\gamma^{\mu}
    +f_6^{b}p_2^{\mu}\gamma^{\nu}
    +f_7^{b}p_1^{\mu}p_1^{\nu}\nonumber\\
&  &  +f_8^{b}p_1^{\mu}p_2^{\nu}
    +f_9^{b}p_2^{\mu}p_1^{\nu}
    +f_{10}^{b}p_2^{\mu}p_2^{\nu}       
    +f_{11}^{b}g^{\mu\nu}(\rlap/p_4)
    +f_{12}^{b}{\rlap/p_4}\gamma^{\mu}\gamma^{\nu}\nonumber\\
& & +f_{13}^{b}{\rlap/p_4}p_1^{\nu}\gamma^{\mu}
    +f_{14}^{b}{\rlap/p_4}p_1^{\mu}\gamma^{\nu}
    +f_{15}^{b}{\rlap/p_4}p_2^{\nu}\gamma^{\mu}
    +f_{16}^{b}{\rlap/p_4}p_2^{\mu}\gamma^{\nu}\nonumber\\
& & \left.+f_{17}^{b}{\rlap/p_4}p_1^{\mu}p_1^{\nu}
    +f_{18}^{b}{\rlap/p_4}p_1^{\mu}p_2^{\nu}
    +f_{19}^{b}{\rlap/p_4}p_2^{\mu}p_1^{\nu}
    +f_{20}^{b}{\rlap/p_4}p_2^{\mu}p_2^{\nu}\right ] v(p_1)\\
\delta M^{\Delta (t)}& =& ie^2Q_t^2\frac{\alpha m_t^2}{16\pi m_W^2 s_W^2}
    \epsilon_{\mu}(p_4)\epsilon_{\nu}(p_3)\bar u(p_2)~\left [
     f_1^{\Delta }g^{\mu\nu}
    +f_7^{\Delta }p_1^{\mu}p_1^{\nu}\right.\nonumber\\
&  &\left. +f_8^{\Delta }p_1^{\mu}p_2^{\nu}
    +f_9^{\Delta }p_2^{\mu}p_1^{\nu}
    +f_{10}^{\Delta }p_2^{\mu}p_2^{\nu}\right ] v(p_1)
\end{eqnarray}

The form factors $f_i^s,f_i^v,f_i^{b}$  are presented
in Appendix A.

    The corresponding amplitude squared can be written as
\begin{eqnarray}
\overline{\sum}\vert M_{ren} \vert^2&=&\overline{\sum}\vert M_0 \vert^2
+2Re\overline{\sum}( \delta M^{(t)} M_0^{(t)^\dagger}
    +\delta M^{(t)} M_0^{(u)^\dagger}
    +\delta M^{(u)} M_0^{(t)^\dagger}\nonumber \\
 & &    +\delta M^{(u)} M_0^{(u)^\dagger})
\end{eqnarray}

$\overline{\sum}\delta M^{(t)} M_0^{(t)^\dagger}$ is given by
\begin{eqnarray}
\overline{\sum}\delta M^{(t)} M_0^{(t)^\dagger}
&=&\delta M^{s(t)}M_0^{(t)^\dagger}+\delta M^{v(t)}M_0^{(t)^\dagger}
  +\delta M^{b(t)}M_0^{(t)^\dagger}
  +\delta M^{\Delta(t)}M_0^{(t)^\dagger},
\end{eqnarray}
where

\begin{eqnarray}
\overline{\sum}\delta M^{s(t)}M_0^{(t)^\dagger}&=&\frac{\pi\alpha^3m_t^2Q_t^2}
    {4m_W^2s_W^2(\hat{t}-m_t^2)}(f_2^{s}H_2+f_6^{s}H_6+f_{12}^{s}H_{12}),\\
\overline{\sum}\delta M^{v(t)}M_0^{(t)^\dagger}&=&\frac{\pi\alpha^3m_t^2Q_t^2}
    {4m_W^2s_W^2(\hat{t}-m_t^2)}(f_2^{v}H_2
    +f_3^{v}H_3+f_6^{v}H_6\nonumber\\
 & &+f_{12}^{v}H_{12}+f_{13}^{v}H_{13}
    +f_{16}^{v}H_{16}),\\
\overline{\sum}\delta M^{b(t)}M_0^{(t)^\dagger}&=&\frac{\pi\alpha^3m_t^2Q_t^2}
    {4m_W^2s_W^2(\hat{t}-m_t^2)}\sum_{i=1}^{20}f_i^{b}H_i
    (m_t,p_1\cdot p_2,p_1\cdot p_3,p_1\cdot p_4,p_3\cdot p_4,),\\
\overline{\sum}\delta M^{\Delta(t)}M_0^{(t)^\dagger}\!&=&\!\frac{\pi
    \alpha^3m_t^2Q_t^2}{4m_W^2s_W^2(\hat{t}-m_t^2)}(f_1^{\Delta}H_1
    +f_{7}^{\Delta}H_{7}+f_{8}^{\Delta}H_{8} 
    +f_{9}^{\Delta}H_{9}+f_{10}^{\Delta}H_{10}),
\end{eqnarray}
Here the expressions of the $H_i(m_t,~p_1\cdot p_2,~p_1\cdot p_3,~p_1\cdot p_4,~
p_3\cdot p_4)$ are given in Appendix B.
$\delta M^{(t)} M_0^{(u)\dagger},\delta M^{(u)} M_0^{(t)\dagger}$ and
$\delta M^{(u)} M_0^{(u)\dagger}$ can be obtained by

\begin{eqnarray}
\delta M^{(t)}{\it M_0}^{(u)^\dagger} 
& = & \delta M^{(t)}{\it M_0}^{(t)\dagger}\vert_{p_3\rightarrow p_4}\\ 
\delta M^{(u)}{\it M_0}^{(t)^\dagger} 
& = & \delta M^{(t)}{\it M_0}^{(t)^\dagger}\vert_{p_4\rightarrow p_3}\\ 
\delta M^{(u)}{\it M_0}^{(u)^\dagger}
& = & \delta M^{(t)}{\it M_0}^{(t)^\dagger}\vert_{p_4\leftrightarrow p_3}
\end{eqnarray}

     The cross section of the subprocess is given by
\begin{equation}
\hat{\sigma(s)}=\frac{1}{16\pi\hat{s}^2}\int_{\hat{t}^-}^{\hat{t}^+}
d\hat{t}\vert M_{ren}(\hat{s},\hat{t}) \vert^2,
\end{equation}
where $~ \hat{t}^\pm=(m_t^2-\frac{1}{2}\hat{s})\pm\frac{1}{2}\hat{s}\beta_t~ $
and $ \beta_t=\sqrt{1-4m_t^2/\hat{s}} $ . The total cross section for
top quark pair production can be obtained by folding the cross section
$ \hat{\sigma}$ for the subprocesses with the photon luminosity
\begin{equation}
\sigma(s)=\int_{2m_t/\sqrt{\hat{s}}}^{x_{max}}dzdL_{\gamma\gamma}/
dz\hat{\sigma}(\gamma\gamma\rightarrow t \bar{t}~{\rm at}~\hat{s}=z^2s)
\end{equation}
where $ \sqrt{s}(\sqrt{\hat{s}}) $ is the $ e^+e^-(\gamma\gamma) $
center-of-mass energy and the quantity $ \frac{dL_{\gamma\gamma}}{dz} $
is the photon luminosity, which is defined as
\begin{equation}
\frac{dL_{\gamma\gamma}}{dz}=2z \int_{\frac{z^2}{x_{max}}}^{x_{max}}
\frac{dx}{x}F_{\gamma/e}(x)F_{\gamma/e}(z^2/x)
\end{equation}
For unpolarized initial electrons and laser, the energy spectrum of the
back-scattered photon is given by [12]
\begin{equation}
F_{\gamma/e}(x)=\frac{1}{D(\xi)}\left[1-x+\frac{1}{1-x}-\frac{4x}{\xi(1-x)}
+\frac{4x^2}{\xi^2(1-x)^2}\right]
\end{equation}
where
\begin{equation}
D(\xi)=(1-\frac{4}{\xi}-\frac{8}{\xi^2})\ln(1+\xi)+\frac{1}{2}
+\frac{8}{\xi}-\frac{1}{2(1+\xi)^2}
\end{equation}
and $ \xi=4E_0\omega_0/m_e^2 $ , $ m_e $ and $ E_0 $ are the incident
electron mass and energy, respectively, and $ \omega_0 $ is the laser-
photon energy, $ x $ is the fraction of the energy of the incident electron
carried by the back-scattered photon. In our calculation we follow the
analysis of Ref. [9], and choose $ \omega_0 $ such that it maximizes
the back-scattered photon energy without spoiling the luminosity through
 $ e^+e^- $ pair creation. With this choice, we can finds $ \xi=2(1+\sqrt{2})
\simeq 4.8 $, $ x_{max}\simeq 0.83 $, and $ D(\xi)\simeq 1.8 $ .

\begin{center}{\large 3. Numerical results and conclusion} \end{center}

In our numerical calculation, the input parameters[13] are $ m_Z=91.176
$GeV,$~\alpha_{em}=1/128.8 $, and $G_F=1.166372 (10^{-5}({\rm GeV})^{-2})$.
$m_W$ is determined through $[4]$
\begin{equation}
m_W^2(1-\frac{m_W^2}{M_Z^2})=\frac{\pi\alpha}{\sqrt 2 G_F}\frac{1}{1-\Delta r},
\end{equation}
where, to order $O(\alpha m_t^2/m_W^2)$, $\Delta r$ is given by $[4]$
\begin{equation}
\Delta r\sim -\frac{\alpha N_Cc_W^2m_t^2}{16\pi^2s_W^4m_W^2}.
\end{equation}
 The lower limit of the parameter $\tan\beta$ is 0.6 from perturbative 
 bounds [14]. Reference [15] argues lower values of $\tan\beta$ from
perturbative unitarity which is about 0.25 for top quark mass of
176 GeV. So in our numerical calculations we allow $\tan\beta$ to take
the minimum value of 0.25 in the two Higgs doublet model.  
   In the following we present some numerical 
examples corresponding to a $e^+e^-$ collider with  center-of-mass energy of 
$\sqrt s=500$ GeV.  

 The numerical results in the SM are presented in Fig.2.
 The correction to the cross section 
depends on the Higgs mass and at $M_h=300$ GeV the correction reaches
its maximum size of $-2.7$\%. Recently, 
the correction in the standard model has been calculated in Ref.[16]. 
But in that work the authors only present
the correction to subprocess cross section $\gamma\gamma\rightarrow t\bar t$ 
and did not give the corresponding results at $e^+e^-$ collider. So 
it is difficult to compare their results with ours. 
   
 We present the numerical results in the two-Higgs-doublet model in  
 Fig.3 and Fig.4. 
In our results we fix the parameters $\alpha$ and $\beta$ to be
$\alpha=\beta=0.25$ and show the dependence on the masses of Higgs bosons.
 The correction is sensitive to the Higgs masses and can be quite large
 for small Higgs masses.
 Fig.3 shows the dependence on the mass of CP-even Higgs bosons 
 $h$ and $H$ for fixed $M_A$ value. 
  We found that correction can be quite large for small $M_h$ value.
  For $M_h<100$ GeV the correction can exceed 50\% and makes it necessary
  to calculate higher order corrections beyond one-loop level. 
  Fig.4 shows the dependence on the mass of CP-odd Higgs boson $A$ for 
  fixed $M_{h,H}$ value. 
 For $M_A=100$ GeV the correction reaches -38\% and decreases rapidly with
 the increase of $M_A$. The corrections drops rapidly with the increase
of $\tan\beta$ as the case of the minimal supersymmetric model discussed
bellow. Here we did not present the numerical results corresponding to
large $\tan\beta$.    

Figs.5-7 represent some numerical results in the minimal supersymmetric model.
The Higgs sector of the minimal supersymmetric model is a special case
of the two-Higgs-doublet model. In this model the masses and couplings of
the Higgs bosons are controlled by two parameters at tree level, which can be 
taken to be $M_A$ and $\beta$ for example. In our numerical results presented
in Figs.3-5, we show the dependence on $M_A$ for three different values
of $\tan\beta$. From these figures one can find that the correction depends
strongly on the values of $\tan\beta$. The correction is more significant
for smaller $\tan\beta$ values. 
 And for a fixed $\tan\beta$ value the correction can be either positive or 
 negative, depending on the Higgs mass $M_A$.
 For minimum $\beta$ value $\beta=0.25$, the 
 correction gets its positive maximum size of 13\% at $M_A=420$ GeV
and negative maximum  size of -54\% at $M_A=300$ GeV. For $\tan\beta=1$
the positive and negative maximum size of the correction can only reach 
7\% and -1.6\%, respectively. For larger $\tan\beta$ value $\tan\beta=5$,
the behaviour of plot in Fig.7 is different from small $\tan\beta$ plots
in Figs.5,6 since the effect of the coupling $\sim m_b \tan\beta$ becomes
significant when $\tan\beta$ gets large and cancel to some extent the effect 
of the coupling $m_t \cot\beta$.     

In conclusion, we calculate the $O(\alpha m_t^2/m_W^2)$ Yukawa corrections
 to top pair production in photon-photon collision in the standard model (SM),
the two-Higgs-doublet model as well as the minimal 
supersymmetric model.  We found that the correction to the
cross section can only reach a few percent in the SM, but can be quite
significant ($>$10\%) in the 2HDM and MSSM for favorable parameter values.
So these corrections are potentially observable at next generation linear
collider, and thus could be used to set limits on the parameters
of these new models, and the precision study for top pair production in
photon-photon collision at NLC will be a powerful indirect probe for new 
physics beyond the standard model. 

This work was supported by the National Natural Science Foundation 
of China and a grant from the State Commission of Science and
Technology of China. 
\newpage
\vspace{0.2in}
	\begin{center} {\Large Appendix A }\end{center}
\vspace{0.1in}
    The form factors $f_i^s,f_i^v$ are given by
\begin{eqnarray*}
f_2^s & = & \frac{m_t}{m_t^2-\hat{t}}\left\{\sum_{i=H^0,h}\eta_i
	\left[(F_1-F_0)(\hat{t},m_t,m_i)+(F_0-F_1)(m_t^2,m_t,m_i)\right]\right.\\
& &+\sum_{i=A,Z}\eta_i\left[(F_1+F_0)(\hat{t},m_t,m_i)
   -(F_1+F_0)(m_t^2,m_t,m_i)\right]\\
& &+\left.\sum_{i=H^+,W^+}\eta_i\left[F_1(\hat{t},0,m_i)-F_1(m_t^2,0,m_i)
   \right]\right\}\\
f_6^s & = &2f_{12}^s= -\frac{4}{(m_t^2-\hat{t})^2}\left \{
    \sum_{i=H^0,h}\eta_i\left[(m_t^2-p_1\cdot p_3)\left(
    F_1(\hat{t},m_t,m_i)-F_1(m_t^2,m_t,m_i)\right)\right.\right.\\ 
& & \left.+2m_t^2p_1\cdot p_3(G_0+G_1)(m_t^2,m_t,m_i)-m_t^2
    \left(F_0(\hat{t},m_t,m_i)-F_0(m_t^2,m_t,m_i)\right)\right] \\ 
& &+\sum_{i=A,Z}\eta_i\left[(m_t^2-p_1\cdot p_3)\left(
   F_1(\hat{t},m_t,m_i)-F_1(m_t^2,m_t,m_i)\right)\right.\\
& &\left.+2m_t^2p_1\cdot p_3(G_1-G_0)(m_t^2,m_t,m_i)-m_t^2
   \left(F_0(m_t^2,m_t,m_i)-F_0(\hat{t},m_t,m_i)\right)\right] \\
& &+\sum_{i=H^+,W^+}\eta_i\left[(m_t^2-p_1\cdot p_3)\left(
    F_1(\hat{t},0,m_i)-F_1(m_t^2,0,m_i)\right)
   +2m_t^2p_1\cdot p_3G_1(m_t^2,0,m_i)\right]\Bigg\} \\
f_2^v & = & \frac{m_t}{\hat{t}-m_t^2}\left\{ 2\sum_{i=A,Z}\eta_i
\left[ p_2\cdot p_4C_{11}
(-p_2,p_4,m_i,m_t,m_t)\right.\right.\\
& &+\left.\left(m_t^2C_{21}+p_1\cdot p_3(C_{11}-2C_{12}-2C_{23})
+2\overline{C}_{24}\right)(p_1,-p_3,m_i,m_t,m_t)\right]\\
& &+2\sum_{i=H^0,h}\eta_i \left[ p_2\cdot p_4(2C_0+C_{11})
(-p_2,p_4,m_i,m_t,m_t)\right.\\
& &+\left.\left(m_t^2(C_{21}-4C_0)+p_1\cdot p_3(2C_0+C_{11}-2C_{12}-2C_{23})
+2\overline{C}_{24}\right)(p_1,-p_3,m_i,m_t,m_t)\right]\\
& &-\sum_{i=H^+,W^+}\eta_i \left[ \left(m_t^2(C_0+C_{21})
+p_1\cdot p_3(C_{11}-2C_{23})+2\overline{C}_{24}\right)(p_1,-p_3,m_i,m_b,m_b)\right.\\
& & \left.-6\overline{C}_{24}(p_1,-p_3,m_b,m_i,m_i)
+p_2\cdot p_4(C_0+C_{11})(-p_2,p_4,m_i,m_b,m_b)\right]\Bigg\}\\
f_3^v & = & \frac{1}{\hat{t}-m_t^2}\left\{4\sum_{i=A,Z}\eta_i \left[m_t^2C_{21}-p_1\cdot p_3
(C_{12}+C_{23})\right]\right.(p_1,-p_3,m_i,m_t,m_t)\\
& &+4\sum_{i=H^0,h}\eta_i \left[m_t^2(2C_{11}+C_{21})-p_1\cdot p_3
(C_{12}+C_{23})\right](p_1,-p_3,m_i,m_t,m_t)\\
& & +2 \sum_{i=H^+,W^+} \eta_i \left\{ \left[-m_t^2(C_0+C_{11}+C_{21})
+p_1\cdot p_3(C_{12}+C_{23})\right]\right.(p_1,-p_3,m_i,m_b,m_b)\\
& &+3\left.\left[-m_t^2(C_{11}+C_{21})+p_1\cdot p_3(C_{12}+C_{23})\right]
(p_1,-p_3,m_b,m_i,m_i)\right\}\Bigg\}\\
f_6^v & = & \frac{1}{\hat{t}-m_t^2}\left\{ 2\sum_{i=A,Z}\eta_i (2\overline{C}_{24}
-m_t^2C_{21})(-p_2,p_4,m_i,m_t,m_t)\right. \\
& &+2\sum_{i=H^0,h}\eta_i \left[2\overline{C}_{24}-m_t^2(4C_0+4C_{11}+C_{21}\right]
(-p_2,p_4,m_i,m_t,m_t) \\ 
& &+\sum_{i=H^+,W^+}\eta_i \left\{ \left[ m_t^2(C_0+C_{11}+C_{21})
-2\overline{C}_{24}\right] \right. (-p_2,p_4,m_i,m_b,m_b)\\
& &+3\left.\left[ 2m_t^2(C_{11}+C_{21})-2p_2\cdot p_4(C_{12}+C_{23})+2\overline{C}_{24}
\right](-p_2,p_4,m_b,m_i,m_i)\right\}\Bigg\} \\
f_{12}^v & = & \frac{1}{\hat{t}-m_t^2}\left\{\sum_{i=A,Z}\eta_i 
\left\{ \left[2\overline{C}_{24}+m_t^2C_{21}
-2p_2\cdot p_4(C_{12}+C_{23})\right](-p_2,p_4,m_i,m_t,m_t)\right.\right.\\
& &+\left.\left[2\overline{C}_{24}+m_t^2C_{21}-2p_1\cdot p_3(C_{12}+C_{23})
\right](p_1,-p_3,m_i,m_t,m_t)\right\}\\
& &+\sum_{i=H^0,h}\eta_i \left\{ \left[2\overline{C}_{24}+m_t^2(C_{21}-4C_0)-2p_2\cdot p_4
(C_{12}+C_{23})\right](-p_2,p_4,m_i,m_t,m_t)\right.\\
& &+\left.\left[m_t^2(C_{21}-4C_0)-2p_1\cdot p_3(C_{12}+C_{23})
+2\overline{C}_{24}\right](p_1,-p_3,m_i,m_t,m_t)\right\}\\
& &+\sum_{i=H^+,W^+}\eta_i \left\{ \left[p_2\cdot p_4(C_{12}+C_{23})
+\frac{1}{2}m_t^2(C_0-C_{21})-\overline{C}_{24}\right](-p_2,p_4,m_i,m_b,m_b)\right.\\
& &+\left[p_1 \cdot p_3(C_{12}+C_{23})-\frac{1}{2}m_t^2(C_0+C_{21})-\overline{C}_{24}
\right](p_1,-p_3,m_i,m_b,m_b)\\
& &+\left.3\overline{C}_{24}(-p_2,p_4,m_b,m_i,m_i)
+3\overline{C}_{24}(p_1,-p_3,m_b,m_i,m_i)\right\}\Bigg\}\\
f_{13}^v & = & \frac{m_t}{\hat{t}-m_t^2}\left\{2\sum_{i=A,Z}\eta_i
C_{21}(p_1,-p_3,m_i,m_t,m_t)\right.\\
& &+2\sum_{i=H^0,h}\eta_i (2C_{11}+C_{21})(p_1,-p_3,m_i,m_t,m_t)\\
& &-\sum_{i=H^+,W^+}\eta_i \left[(C_0+C_{11}+C_{21})(p_1,-p_3,m_i,m_b,m_b)
\right.\\
& &\left.+3(C_{11}+C_{21})(p_1,-p_3,m_b,m_i,m_i)\right]\Bigg\}\\
f_{16}^v & = & \frac{m_t}{\hat{t}-m_t^2}\left\{2\sum_{i=A,Z}\eta_i
C_{21}(-p_2,p_4,m_i,m_t,m_t)\right.\\
& &+2\sum_{i=H^0,h}\eta_i (2C_{11}+C_{21})(-p_2,p_4,m_i,m_t,m_t)\\
& &-\left. \sum_{i=H^+,W^+}\eta_i \left[(C_{11}+C_{21})(-p_2,p_4,m_i,m_b,m_b)
+3(C_{11}+C_{21})(-p_2,p_4,m_b,m_i,m_i)\right]\right\}\\
\end{eqnarray*}
where
\begin{eqnarray*}
F_n(q,m_1,m_2) & = &\int_0^1 dy\,y^n\log\left[\frac{-q^2y(1-y)+m_1^2(1-y)+m_2^2y}{\mu^2}\right],\\
G_n(q,m_1,m_2) & = &-\int_0^1 dy\frac{y^{n+1}(1-y)}{-q^2y(1-y)+m_1^2(1-y)+m_2^2y},
\end{eqnarray*} 
 and $\overline{C}_{24}\equiv -\frac{1}{4}\Delta + C_{24}, C_0, C_{ij}$
are the three-point Feynman integrals, definition and expression for
which can be found in Ref.[17].\par 
The form factor  $f_i^b$ are given by
 $$
 f_i^{b}=f_i^{b(1)}+f_i^{b(2)}+f_i^{b(3)}-f_i^{b(4)}+f_i^{b(5)}
 $$
where
\begin{eqnarray*}
f_1^{b(1)} & = & 2m_t\sum_{i=H^0,h}\eta_i (2D_{27}-D_{311}+2D_{312})\\
f_2^{b(1)} & = & m_t\sum_{i=H^0,h}\eta_i \left[m_t^2(D_0-D_{11}+D_{12}-D_{21}
-D_{22}+2D_{24}+D_{31}-D_{32}\right.\\
& &-D_{34}+D_{36})+2p_3\cdot p_4(D_{13}-D_{12}-D_{310}+D_{26})\\
& &+2(2D_{311}-3D_{312}-2D_{27})+2p_2\cdot p_4(D_{21}-2D_{24}-D_{25}+D_{26}
+D_{38}-D_0)\\
& &+2p_1\cdot p_4(D_{22}-D_{35})+2p_1\cdot p_2(D_{34}-D_{36})\Big]\\
f_3^{b(1)} & = & \sum_{i=H^0,h}\eta_i \left[m_t^2(D_{13}+2D_{35}+2D_{38}
-4D_{39})-4p_2\cdot p_4(D_{12}-D_{13}+D_{23}\right.\\
& &+D_{24}-D_{25}+D_{39})-4p_2\cdot p_3(D_{23}+D_{37}+D_{310})\\
& &+4p_3\cdot p_4D_{26}+4(D_{27}-D_{312}+3D_{313})\Big]\\
f_4^{b(1)} & = & 2\sum_{i=H^0,h}\eta_i \left[m_t^2(-3D_{12}+3D_{13}+2D_{23}
+2D_{24}-2D_{25}-2D_{26}+D_{32}+D_{34}\right.\\
& &-D_{35}-D_{38})+2p_2\cdot p_4(D_{12}-D_{13}+D_{24}-D_{25}
-D_{38}+D_{39})\\
& &+2p_1\cdot p_2(D_{22}+D_{23}-2D_{26}+D_{36}-D_{310})\\
& &+2p_2\cdot p_3D_{37}+4(D_{312}-D_{313})\Big]\\
f_5^{b(1)} & = & 2\sum_{i=H^0,h}\eta_i \left[m_t^2(-D_{11}+D_{21}+D_{22}
+D_{31}+D_{36}-D_0)\right.\\
& &+2p_1\cdot p_4(D_{12}+D_{13}+D_{24}-2D_{25}-D_{35})
-2p_1\cdot p_3(D_{26}+D_{310})\\
& &+2p_1\cdot p_2(D_{24}+D_{34})+2(2D_{27}+D_{311})\Big]\\
f_6^{b(1)} & = & 2\sum_{i=H^0,h}\eta_i \left[m_t^2(-2D_{11}+2D_{13}
+D_{21}-D_{22}+2D_{23}+D_0+2D_{26}-2D_{25}\right.\\
& &+2D_{39}+2D_{310})+2p_1\cdot p_4(-D_{12}+D_{13}+D_{25})
-2p_2\cdot p_4(D_{23}+D_{310})\\
& &-2p_1\cdot p_2D_{24}-(2D_{27}+D_{311})\Big]\\
f_7^{b(1)} & = & 4m_t\sum_{i=H^0,h}\eta_i (D_{22}-D_{26}+D_{32}-D_{36}
-D_{38}+D_{310})\\
f_8^{b(1)} & = & 4m_t\sum_{i=H^0,h}\eta_i (D_{12}-D_{13}+D_{22}-D_{26}
-D_{34}+D_{35}+D_{36}-D_{310})\\
f_9^{b(1)} & = & 4\sum_{i=H^0,h}\eta_i \left[m_t(D_{24}-D_{26}
-D_{34}+D_{36}-D_{38}+D_{310})+D_{313}\right]\\
f_{10}^{b(1)} & = & 4m_t\sum_{i=H^0,h}\eta_i (D_{11}-D_{13}+D_{24}-D_{26}
-D_{31}+D_{34}+D_{35}-D_{310})\\
f_{11}^{b(1)} & = & -4\sum_{i=H^0,h}\eta_i (D_{27}+D_{313})\\
f_{12}^{b(1)} & = & \sum_{i=H^0,h}\eta_i \left[m_t^2(3D_0-2D_{11}+D_{13}-D_{21}
-D_{22}-2D_{23}+2D_{25}+2D_{26}+D_{35}+D_{38}\right.\\
& &-D_{39}+D_{310})+2(D_{27}+3D_{313})+2p_1\cdot p_2(-D_{12}+D_{13}
-D_{23}-D_{24}+D_{25}\\
& &-D_{26}-D_{310})-2p_1\cdot p_3(D_{39}+D_{310})
-2p_2\cdot p_3D_{37}\Big]\\
f_{13}^{b(1)} & = & -2m_t\sum_{i=H^0,h}\eta_i (D_{12}+D_{22}-D_{24}+2D_{39}
+2D_{310})(-p_2,-p_1,p_3,m_t,m_i,m_t,m_t)\\
f_{14}^{b(1)} & = & 2m_t\sum_{i=H^0,h}\eta_i (D_{22}-D_{24}+D_{25}
-D_{26})\\
f_{15}^{b(1)} & = & -2m_t\sum_{i=H^0,h}\eta_i (D_0+D_{12}-D_{21}+D_{24}+2D_{39}
+2D_{310})\\
f_{16}^{b(1)} & = & 2m_t\sum_{i=H^0,h}\eta_i (D_0+D_{12}-D_{13}-D_{21}+D_{24}
+D_{25}-D_{26})\\
f_{17}^{b(1)} & = & 4\sum_{i=H^0,h}\eta_i (D_{23}-D_{26}-D_{38}+D_{39})\\
f_{18}^{b(1)} & = & 4\sum_{i=H^0,h}\eta_i (-D_{12}+D_{13}+D_{23}-D_{24}+D_{25}
-D_{26}+D_{37}-D_{310})\\
f_{19}^{b(1)} & = & 4\sum_{i=H^0,h}\eta_i (D_{12}-D_{13}+D_{23}+D_{24}-D_{25}
-D_{26}+D_{39}-D_{310})\\
f_{20}^{b(1)} & = & 4\sum_{i=H^0,h}\eta_i (D_{23}-D_{25}-D_{35}+D_{37})\\
f_1^{b(2)} & = & {1\over 2} m_t\sum_{i=H^+,W^+}\eta_i (2D_{312}-D_{311})\\
f_2^{b(2)} & = &{1\over 4} m_t \sum_{i=H^+,W^+}\eta_i \left[m_t^2(D_{31}-D_{32}
-D_{34}+D_{36})\right.\\
& &+2p_2\cdot p_4(D_{11}-D_{12}+D_{21}-D_{24}+D_{38})
+2p_1\cdot p_4(D_{24}-D_{22}-D_{35})\\
& &+2p_1\cdot p_2(D_{34}-D_{36})+2p_3\cdot p_4(D_{26}-D_{25}-D_{310})
+2(2D_{311}-3D_{312})\Big]\\
f_3^{b(2)} & = & {1\over 2} \sum_{i=H^+,W^+}\eta_i \left[m_t^2(D_{35}+D_{38}
-2D_{39}-2D_{310})+2p_3\cdot p_4(D_{26}-D_{23}) \right.\\
& &+2p_2\cdot p_4(D_{13}-D_{12}-D_{24}+D_{25}-D_{39}+D_{310})\\
& &-2p_2\cdot p_3D_{37}-2p_1\cdot p_2D_{310}+2(D_{27}-D_{312}+3D_{313})
\Big]\\
f_4^{b(2)} & = &{1\over 2} \sum_{i=H^+,W^+}\eta_i \left[m_t^2(2D_{24}-2D_{25}
-2D_{26}+D_{32}+D_{34}-D_{35}-D_{38})\right.\\
& &+2p_2\cdot p_4(D_{12}-D_{13}+D_{24}-D_{25}-D_{38}+D_{39})
+2p_3\cdot p_4D_{23}+2p_2\cdot p_3D_{37}\\
& &+2p_1\cdot p_2(D_{22}-2D_{26}+D_{36}-D_{310})
+4(D_{312}-D_{313})\Big]\\
f_5^{b(2)} & = &{1\over 2} \sum_{i=H^+,W^+}\eta_i \left[m_t^2(D_{21}+D_{22}
+D_{31}+D_{36})\right.\\
& &+2p_1\cdot p_2(D_{24}+D_{34})-2p_1\cdot p_3(D_{26}+D_{310})\\
& &+2p_1\cdot p_4(D_{12}-D_{13}+D_{24}-2D_{25}-D_{35})
+2(D_{311}+2D_{27})\Big]\\
f_6^{b(2)} & = &{1\over 2} \sum_{i=H^+,W^+}\eta_i \left[m_t^2(D_{21}-D_{22}
-2D_{25}+2D_{26}+2D_{39}+2D_{310})\right.\\
& &+2p_1\cdot p_4(D_{13}-D_{12}-D_{24}+D_{25})
-2p_1\cdot p_3D_{310}-(2D_{27}+D_{311})\Big]\\
f_7^{b(2)} & = &m_t\sum_{i=H^+,W^+}\eta_i (D_{32}-D_{36}-D_{38}+D_{310})\\
f_8^{b(2)} & = &m_t\sum_{i=H^+,W^+}\eta_i (D_{22}-D_{24}+D_{25}-D_{26}
-D_{34}+D_{35}+D_{36}-D_{310})\\
f_9^{b(2)} & = & \sum_{i=H^+,W^+}\eta_i\left[m_t(D_{36}-D_{34}-D_{38}+D_{310})
+D_{313}\right]\\
f_{10}^{b(2)} & = &m_t\sum_{i=H^+,W^+}\eta_i (-D_{21}+D_{24}+D_{25}-D_{26}
-D_{31}+D_{34}+D_{35}-D_{310})\\
\end{eqnarray*}
\begin{eqnarray*}
f_{11}^{b(2)} & = &-\sum_{i=H^+,W^+}\eta_i(D_{27}+D_{313})\\
f_{12}^{b(2)} & = &{1\over 4} \sum_{i=H^+,W^+}\eta_i \left[m_t^2(-2D_{11}+2D_{13}
-D_{21}-D_{22}+2D_{25}+2D_{26}-D_{35}+D_{38}\right.\\
& &-D_{39}+D_{310})+2p_1\cdot p_2(-D_{12}+D_{13}-D_{24}+D_{25}-D_{26}-D_{310})
\\
& &-2p_1\cdot p_3(D_{39}+D_{310})-2p_3\cdot p_4D_{23}-2p_2\cdot p_3D_{37}
+2(D_{27}+3D_{313})\Big]\\
f_{13}^{b(2)} & = & {1\over 2}m_t\sum_{i=H^+,W^+}\eta_i (-D_{22}+D_{24}-2D_{39}
-2D_{310})\\
f_{14}^{b(2)} & = & {1\over 2}m_t\sum_{i=H^+,W^+}\eta_i (D_{22}-D_{24}+D_{25}
-D_{26})\\
f_{15}^{b(2)} & = & {1\over 2}m_t\sum_{i=H^+,W^+}\eta_i (D_{11}-D_{12}+D_{21}
-D_{24}-2D_{39}-2D_{310})\\
f_{16}^{b(2)} & = & {1\over 2}m_t\sum_{i=H^+,W^+}\eta_i (-D_{11}+D_{12}-D_{21}
+D_{24}+D_{25}-D_{26})\\
f_{17}^{b(2)} & = & \sum_{i=H^+,W^+}\eta_i (D_{23}-D_{26}
-D_{38}+D_{39})\\
f_{18}^{b(2)} & = & \sum_{i=H^+,W^+}\eta_i (-D_{12}+D_{13}+D_{23}
-D_{24}+D_{25}-D_{26}+D_{37}-D_{310})\\
f_{19}^{b(2)} & = & \sum_{i=H^+,W^+}\eta_i (D_{12}-D_{13}+D_{23}
+D_{24}-D_{25}-D_{26}+D_{39}-D_{310})\\
f_{20}^{b(2)} & = & \sum_{i=H^+,W^+}\eta_i (D_{23}-D_{25}-D_{35}
+D_{37})\\
f_1^{b(3)} & = &-9m_t \sum_{i=H^+,W^+}\eta_i (-D_{27}-D_{311}+D_{312})\\
f_2^{b(3)} & = &0\\
f_3^{b(3)} & = & 9 \sum_{i=H^+,W^+}\eta_i (D_{27}+D_{312})\\
f_4^{b(3)} & = & -9 \sum_{i=H^+,W^+}\eta_i (D_{313}-D_{312})\\
f_5^{b(3)} & = & 9 \sum_{i=H^+,W^+}\eta_i (D_{27}+D_{311})\\
f_6^{b(3)} & = & -9 \sum_{i=H^+,W^+}\eta_i (D_{313}-D_{311})\\
f_7^{b(3)} & = & -9 m_t\sum_{i=H^+,W^+}\eta_i (D_{13}-D_{12}-D_{24}
+D_{25}+D_{32}-D_{36}-D_{38}+D_{310})\\
f_8^{b(3)} & = & -9 m_t\sum_{i=H^+,W^+}\eta_i (D_{13}-D_{12}-2D_{24}
+D_{25}+D_{22}\\
& &-D_{26}-D_{34}+D_{35}+D_{36}-D_{310})\\
f_9^{b(3)} & = & -9 m_t\sum_{i=H^+,W^+}\eta_i (D_{13}-D_{11}-D_{21}
+D_{25}-2D_{34}+D_{36}-D_{38}+D_{310})\\
f_{10}^{b(3)} & = & -9 m_t\sum_{i=H^+,W^+}\eta_i (D_{13}-D_{11}-2D_{21}
+2D_{25}+D_{24}\\
& &-D_{26}-D_{31}+D_{35}-D_{310})\\
f_{11}^{b(3)} & = & 9\sum_{i=H^+,W^+}\eta_i D_{313}\\
f_{12}^{b(3)}&=&f_{13}^{b(3)}=f_{14}^{b(3)}=f_{15}^{b(3)}=f_{16}^{b(3)}=0,\\
f_{17}^{b(3)} & = & -9 \sum_{i=H^+,W^+}\eta_i (D_{23}-D_{26}+D_{39}
-D_{38})\\
f_{18}^{b(3)} & = & -9 \sum_{i=H^+,W^+}\eta_i (D_{23}-D_{26}+D_{37}
-D_{310})\\
f_{19}^{b(3)} & = & -9 \sum_{i=H^+,W^+}\eta_i (D_{23}-D_{25}+D_{39}
-D_{310})\\
f_{20}^{b(3)} & = & -9 \sum_{i=H^+,W^+}\eta_i (D_{23}-D_{25}+D_{37}
-D_{35})\\
f_1^{b(4)} & = & 2m_t \sum_{i=A,Z} \eta_i (2D_{27}-D_{311}+2D_{312})\\
f_2^{b(4)} & = & m_t \sum_{i=A,Z}\eta_i \left[m_t^2(D_0-D_{11}
+D_{12}+D_{21}+D_{22}+2D_{24}+D_{31}-D_{32}-D_{34}+D_{36})\right.\\
& &+2p_3\cdot p_4(D_{13}-D_{12}-D_{310}+D_{26})
+2p_2\cdot p_4(-D_0+D_{21}-2D_{24}-D_{25}\\
& &-D_{26}+D_{38})+2p_1\cdot p_4(-2D_{13}+D_{22}-2D_{25}-D_{35}+2 D_{12})
+2p_1\cdot p_2(2D_{24}\\
& &+D_{34}-D_{36})+2(2D_{311}-3D_{312})\Big]\\
f_3^{b(4)} & = & \sum_{i=A,Z} \eta_i \left[m_t^2(D_{13}+2D_{35}
+2D_{38}-4D_{39})\right.\\
& &-4p_2\cdot p_4(D_{12}-D_{13}+D_{23}+D_{24}-D_{25}+D_{39})
+4p_3\cdot p_4D_{26}\\
& &-4p_2\cdot p_3(D_{23}+D_{37}+D_{310})
+4(D_{27}-D_{312}+3D_{313})\Big]\\
f_4^{b(4)} & = & 2 \sum_{i=A,Z} \eta_i \left[m_t^2(-D_{12}+D_{13}
-2D_{22}-2D_{23}+4D_{26}+D_{32}+D_{34}-D_{35}-D_{38})\right.\\
& &+2p_2\cdot p_4(D_{12}-D_{13}+D_{24}-D_{25}-D_{38}+D_{39})
+2p_1\cdot p_2(D_{22}+D_{23}\\
& &-2D_{26}+D_{36}-D_{310})+2p_2\cdot p_3D_{37}
+4(D_{312}-D_{313})\Big]\\
\end{eqnarray*}
\begin{eqnarray*}
f_5^{b(4)} & = & 2 \sum_{i=A,Z} \eta_i \left[m_t^2(D_0-D_{11}
+D_{21}+D_{22}+D_{31}+D_{36})\right.\\
& &+2p_1\cdot p_4(D_{12}+D_{13}+D_{24}-2D_{25}-D_{35})
+2p_1\cdot p_2(D_{24}+D_{34})\\
& &-2p_1\cdot p_3(D_{26}+D_{310})+2(2D_{27}+D_{311})\Big]\\
f_6^{b(4)} & = & 2 \sum_{i=A,Z} \eta_i \left[m_t^2(-D_0-D_{21}
-D_{22}-2D_{23}-2D_{24}+4D_{25}\right.\\
& &+4D_{26}+2D_{39}+2D_{310})+2p_1\cdot p_4(-D_{12}+D_{13}+D_{25})
-2p_2\cdot p_4(D_{23}+D_{310})\\
& &-2p_1\cdot p_2D_{24}-(D_{311}+2D_{27})\Big]\\
f_7^{b(4)} & = & 4m_t \sum_{i=A,Z} \eta_i (-D_{22}-2D_{23}+3D_{26}+D_{32}
-D_{36}-D_{38}+D_{310})\\
f_8^{b(4)} & = & 4m_t \sum_{i=A,Z} \eta_i (D_{12}-D_{13}+D_{22}-D_{26}-D_{34}
+D_{35}+D_{36}-D_{310})\\
f_9^{b(4)} & = & 4 \sum_{i=A,Z} \eta_i \left[m_t(-2D_{23}-D_{24}+2D_{25}+D_{26}
-D_{34}+D_{36}\right.\\
& &\left.-D_{38}+D_{310})+D_{313}\right]\\
f_{10}^{b(4)} & = & 4m_t \sum_{i=A,Z} \eta_i (D_{11}-D_{13}+D_{24}-D_{26}
-D_{31}+D_{34}+D_{35}-D_{310})\\
f_{11}^{b(4)} & = & -4 \sum_{i=A,Z} \eta_i (D_{27}+D_{313})\\
f_{12}^{b(4)} & = & \sum_{i=A,Z}\eta_i \left[m_t^2(-3D_{13}-D_{21}
-D_{22}-2D_{23}+2D_{25}+2D_{26}+D_{35}+D_{38}\right.\\
& &-D_{39}+D_{310})-2p_1\cdot p_3(D_{39}+D_{310})
-2p_2\cdot p_3D_{37}\\
& &+2p_1\cdot p_2(-D_{12}+D_{13}-D_{23}-D_{24}+D_{25}-D_{26}-D_{310})\\
& &+2(D_{27}+3D_{313})\Big]\\
f_{13}^{b(4)} & = & -2m_t \sum_{i=A,Z} \eta_i (-D_{12}+2D_{13}+D_{22}-D_{24}
+2D_{39}+2D_{310})\\
f_{14}^{b(4)} & = & 2m_t \sum_{i=A,Z} \eta_i (-2D_{12}+2D_{13}+D_{22}+2D_{23}
-D_{24}+D_{25}-3D_{26})\\
f_{15}^{b(4)} & = & -2m_t \sum_{i=A,Z} \eta_i (D_0+D_{12}-D_{21}+D_{24}
+2D_{39}+2D_{310})\\
f_{16}^{b(4)} & = & 2m_t \sum_{i=A,Z} \eta_i (D_0+D_{12}-D_{13}-D_{21}+2D_{23}
+D_{24}-D_{25}-D_{26})\\
f_{17}^{b(4)} & = & 4 \sum_{i=A,Z} \eta_i (D_{23}-D_{26}-D_{38}+D_{39})\\
f_{18}^{b(4)} & = & 4 \sum_{i=A,Z} \eta_i (-D_{12}+D_{13}+D_{23}-D_{24}+D_{25}
-D_{26}+D_{37}-D_{310})\\
f_{19}^{b(4)} & = & 4 \sum_{i=A,Z} \eta_i (D_{12}-D_{13}+D_{23}+D_{24}-D_{25}
-D_{26}+D_{39}-D_{310})\\
f_{20}^{b(4)} & = & 4 \sum_{i=A,Z} \eta_i (D_{23}-D_{25}-D_{35}+D_{37})\\
f_{1}^{b(5)} & = & \sum_{i=H^+,W^+}\eta_i 2m_tD_{312},f_{2}^{b(5)} =  0,\\
f_{3}^{b(5)} & = & \sum_{i=H^+,W^+}\eta_i m_t^2D_{22}-2D_{27}-2p_2
\cdot p_4(D_{24}+D_{12}-D_{13}+D_{34})\\
& & +2p_3\cdot p_4(D_{25}+D_{35})-p_2\cdot p_3(2D_{26}+D_{310})+4D_{311}
    +m_t^2D_{36}\\    
f_{4}^{b(5)} & = &- \sum_{i=H^+,W^+}\eta_i 2D_{313},\\
f_{5}^{b(5)} & = &\sum_{i=H^+,W^+}\eta_i 4(D_{311}-D_{312})+2p_3\cdot p_4D_{35}-2p_2\cdot p_4D_{34}
-p_2\cdot p_3D_{310}\\
& & +m_t^2D_{36}+2p_2\cdot p_4D_{36}+2p_2\cdot p_3D_{38}-p_3\cdot p_4D_{310}\\
& & -m_t^2D_{32}+2p_2\cdot p_4(D_{22}+D_{25}-D_{24}-D_{26})\\
f_{6}^{b(5)} & = &\sum_{i=H^+,W^+}\eta_i 2(D_{312}-D_{313}),\\
f_{7}^{b(5)} & = &\sum_{i=H^+,W^+}\eta_i 2m_t(D_{26}+D_{310}),\\
f_{8}^{b(5)} & = &\sum_{i=H^+,W^+}\eta_i 2m_t(D_{310}-D_{38}),\\
f_{9}^{b(5)} & = &\sum_{i=H^+,W^+}\eta_i 2m_t(D_{26}-D_{22}+D_{310}-D_{36}),\\
f_{10}^{b(5)} & = &\sum_{i=H^+,W^+}\eta_i 2m_t(D_{32}-D_{38}-D_{310}-D_{36}),\\
f_{11}^{b(5)} & = &\sum_{i=H^+,W^+}\eta_i 2(D_{313}-D_{311}),\\
f_{12}^{b(5)} & = &\sum_{i=H^+,W^+}\eta_i D_{27},\\
f_{13}^{b(5)} & = &\sum_{i=H^+,W^+}\eta_i m_t(D_{12}+D_{24}),\\
f_{14}^{b(5)} & = &0,\\
f_{15}^{b(5)} & = &\sum_{i=H^+,W^+}\eta_i m_t(D_{24}-D_{22}),\\
f_{16}^{b(5)} & = &0,\\
f_{17}^{b(5)} & = &\sum_{i=H^+,W^+}\eta_i 2(D_{23}-D_{25}+D_{37}-D_{35}),\\
f_{18}^{b(5)} & = &\sum_{i=H^+,W^+}\eta_i 2(D_{37}-D_{35}-D_{39}),\\
f_{19}^{b(5)} & = &\sum_{i=H^+,W^+}\eta_i 2(D_{23}+2D_{24}-2D_{25}-D_{26}+D_{12}-D_{13}
   +D_{34}+D_{37}-D_{35}),\\
f_{20}^{b(5)} & = &\sum_{i=H^+,W^+}\eta_i 2(D_{34}+D_{37}+D_{38}-D_{35}-D_{36}
   -D_{39}+D_{24}+D_{26}-D_{22}-D_{25}),\\
f_1^{\Delta }& = &\sum_{i=H^0,h}\frac{\eta_i}{(\hat{s}-m_i^2+im_i\Gamma_i)}
\{-12m_t[m_t^2C_0-p_3\cdot p_4(2C_{22}
-2C_{23}+C_0)+\frac{1}{2}]\}\\
& & (p_4,-p_1-p_2,m_t,m_t,m_t)
+\sum_{i=H^+,W^+}\frac{\eta_i}{Q_t^2}(m_tC_{11})(-p_2,p_1+p_2,m_b,m_i,m_i)\\
f_7^{\Delta}&=&f_8^{\Delta }=f_9^{\Delta}=f_{10}^{\Delta }
 = \sum_{i=H^0,h}\frac{\eta_i}{(\hat{s}-m_i^2+im_i\Gamma_i)}
\{-12m_t[C_0+4(C_{22}-C_{23})]\}\\
& &(p_4,-p_1-p_2,m_t,m_t,m_t)\\
\end{eqnarray*}

In the above, $D_0,D_{ij},D_{ijk}$ are the four-point Feynman
integrals[17], and\\
 $D_0,D_{ij},D_{ijk}(-p_2,-p_1,p_3,m_t,m_i,m_t,m_t)\ \ \ $ in $\;f_i^{b(1)}$ and $\;f_i^{b(4)}$\\
 $D_0,D_{ij},D_{ijk}(-p_2,-p_1,p_3,0,m_i,0,0)\ \ \ \ \ \ \ \ \ \,$ in $\;f_i^{b(2)}$\\
 $D_0,D_{ij},D_{ijk}(-p_2,-p_1,p_3,m_i,0,m_i,m_i)\ \ \ \ \ \,$ in $\;f_i^{b(3)}$\\
 $D_0,D_{ij},D_{ijk}(p_4,-p_2,p_3,m_b,m_b,m_i,m_i)\ \ \ \ \ $ in $\;f_i^{b(5)}$\\
and
$$
\hat{s}=(p_1+p_2)^2,\ \ \ \ \ \hat{t}=(p_3-p_1)^2,\ \ \ \ \ \hat{u}=(p_3-p_2)^2
$$
$$
p_1\cdot p_2=\frac{1}{2}(\hat{s}-2m_t^2),\ \ \ \ \ 
p_1\cdot p_3=p_2\cdot p_4=\frac{1}{2}(m_t^2-\hat{t}),
$$
$$
p_3\cdot p_4=\frac{1}{2}\hat{s},~~~~~p_1\cdot p_4=p_2\cdot p_3=\frac{1}{2}
		(m_t^2-\hat{u})
$$
$$
\eta_{H^0}=\frac{\sin^2\alpha}{\sin^2\beta},\ \ \ \ \ 
\eta_{h}=\frac{\cos^2\alpha}{\sin^2\beta},
$$
$$
\eta_{A}=\cot^2\beta,\;\ \ \ \ \eta_{H^+}=\cot^2\beta+\frac{m^2_b}{m_t^2}
\tan^2\beta,\;\ \ \ \ \eta_{Z}=\eta_{W^+}=1,
$$

\vspace{0.3in}
	\begin{center}{\Large Appendix B }\end{center}
\vspace{0.2in}

    The expressions of $H_i(m_t,p_1\cdot p_2,p_1\cdot p_3,p_1\cdot p_4,p_3\cdot p_4,)$
in the amplitude squared are given by
\begin{eqnarray*}
H_1&=&-8m_t^3+8m_tp_1\cdot p_2-
   8m_tp_1\cdot p_3+8m_tp_2\cdot p_3\\
H_2&=&-32m_t^3-16m_tp_1\cdot p_2-32m_tp_1\cdot p_3+32m_tp_2\cdot p_3\\
H_3&=&32m_t^4+16m_t^2p_1\cdot p_2-16m_t^2p_1\cdot p_3-8m_t^2p_2\cdot p_3\\
H_4&=&8m_t^4-8m_t^2p_1\cdot p_2-16m_t^2p_1\cdot p_3
   -16p_1\cdot p_2 p_1\cdot p_3+8m_t^2p_2\cdot p_3\\
H_5&=&32m_t^2p_1\cdot p_2+16(p_1\cdot p_2)^2-16m_t^2p_2\cdot p_3
   -16p_1\cdot p_2p_2\cdot p_3+8m_t^2p_1\cdot p_3\\
H_6&=&8m_t^2p_1\cdot p_2-8m_t^4-16m_t^2p_2\cdot p_3-8m_t^2p_1\cdot p_3\\
H_7&=&-8m_t^5+8m_t^3p_1\cdot p_2+4m_t^3p_1\cdot p_3-8m_tp_1\cdot p_2
   p_1\cdot p_3+4m_t^3p_2\cdot p_3\\
H_8&=&-8m_t^3p_1\cdot p_2+8m_t(p_1\cdot p_2)^2+4m_t^3p_2\cdot p_3
   - 4m_t^3p_1\cdot p_3\\
H_9&=&-8m_t^3p_1\cdot p_2+8m_t^5+4m_t^3p_2\cdot p_3-4m_t^3p_1\cdot p_3\\
H_{10}&=&-8m_t(p_1\cdot p_2)^2+8m_t^3p_1\cdot p_2+8m_tp_1\cdot p_2p_2\cdot p_3
   -4m_t^3p_1\cdot p_3-4m_t^3p_2\cdot p_3\\
H_{11}&=&8m_t^2p_1\cdot p_4-8m_t^2p_2\cdot p_4+8m_t^2p_3\cdot p_4
   -8p_1\cdot p_4p_2\cdot p_3\\
   & &+8p_1\cdot p_2p_3\cdot p_4-8p_1\cdot p_3p_2\cdot p_4\\
H_{12}&=&-16m_t^2p_1\cdot p_4-32m_t^2p_2\cdot p_4+32m_t^2p_3\cdot p_4
   -32p_1\cdot p_3p_2\cdot p_4\\
H_{13}&=&16m_t^3p_1\cdot p_4+32m_t^3p_2\cdot p_4-8m_t^3p_3\cdot p_4
   -16m_tp_1\cdot p_3p_2\cdot p_4\\
H_{14}&=&-8m_t^3p_1\cdot p_4+8m_t^3p_2\cdot p_4-16m_tp_1\cdot p_3p_1\cdot p_4
   +8m_t^3p_3\cdot p_4\\
   & &-16m_tp_1\cdot p_2p_3\cdot p_4
   +16m_tp_1\cdot p_4p_2\cdot p_3-16m_tp_1\cdot p_3p_2\cdot p_4\\
H_{15}&=&16m_tp_1\cdot p_2p_1\cdot p_4+32m_tp_1\cdot p_2p_2\cdot p_4
   -8m_tp_1\cdot p_4p_2\cdot p_3+8m_tp_1\cdot p_3p_2\cdot p_4\\
   & &-8m_tp_1\cdot p_2p_3\cdot p_4-16m_tp_2\cdot p_3p_2\cdot p_4\\
H_{16}&=&-8m_t^3p_2\cdot p_4+8m_t^3p_1\cdot p_4-8m_tp_1\cdot p_3p_2\cdot p_4
   +8m_tp_1\cdot p_2p_3\cdot p_4\\
   & &-8m_tp_1\cdot p_4p_2\cdot p_3-16m_t^3p_3\cdot p_4\\
H_{17}&=&-8m_t^2p_1\cdot p_3p_1\cdot p_4+4m_t^4p_3\cdot p_4
   +4m_t^2p_1\cdot p_3p_2\cdot p_4-4m_t^2p_1\cdot p_4p_2\cdot p_3\\
    & &+4m_t^2p_1\cdot p_2p_3\cdot p_4+8m_t^4p_1\cdot p_4-8m_t^4p_2\cdot p_4\\
H_{18}&=&8m_t^2p_1\cdot p_2p_1\cdot p_4-8m_t^2p_1\cdot p_2p_2\cdot p_4
   -4m_t^2p_1\cdot p_4p_2\cdot p_3+4m_t^2p_1\cdot p_2p_3\cdot p_4\\
    & &-4m_t^2p_1\cdot p_3p_2\cdot p_4+8p_1\cdot p_2p_1\cdot p_4p_2\cdot p_3
    -8p_1\cdot p_2p_1\cdot p_3p_2\cdot p_4+8(p_1\cdot p_2)^2p_3\cdot p_4\\
    & &-16p_1\cdot p_4p_1\cdot p_2p_2\cdot p_3+8m_t^2p_1\cdot p_3p_1\cdot p_4
   +8m_t^2p_2\cdot p_3p_2\cdot p_4-4m_t^4p_3\cdot p_4\\
H_{19}&=&8m_t^4p_2\cdot p_4-8m_t^4p_1\cdot p_4-4m_t^2p_1\cdot p_3
   p_2\cdot p_4\\
    & &+4m_t^2p_1\cdot p_2p_3\cdot p_4-4m_t^2p_1\cdot p_4p_2\cdot p_3
   +4m_t^4p_3\cdot p_4\\
H_{20}&=&8m_t^2p_1\cdot p_2p_2\cdot p_4-8m_t^2p_1\cdot p_2p_1\cdot p_4
   -8m_t^2p_2\cdot p_3p_2\cdot p_4+4m_t^4p_3\cdot p_4\\
    & &+4m_t^2p_1\cdot p_4p_2\cdot p_3-4m_t^2p_1\cdot p_3p_2\cdot p_4
   +4m_t^2p_1\cdot p_2p_3\cdot p_4\\
\end{eqnarray*}

\baselineskip=0.3in
\vspace{1.2in}
{\LARGE References}
\vspace{0.3in}
\begin{itemize}
\begin{description}

\item[{\rm[1]}] CDF Collaboration, F. Abe {\it et al.}, Phys. Rev. {\bf D50},
2966(1994).
\item[{\rm[2]}] V. Barger and R. J. N. Phillips, Institution Report No.
MAD/PH/780, 1993(unpublished).
\item[{\rm[3]}] A. Blondel, F. M. Renard, and C. Verzegnassi, Phys. lett.
{\bf B269}, 419(1991).
\item[{\rm[4]}] M. E. Peskin, in Physics and Experiments with Linear Colliders,
Proceedings of the Workshop, Saariselk\"a, Finland, 1991, edited by R. Orava,
P. Eerala, and M. Nordberg(World scientific,Singapore, 1992), P.1.
\item[{\rm[5]}] I. F. Ginzbyrg, G. L. Kotkin, V. G. Serbo and V. I. Telnov,
Pis'ma ZHETF 34(1981)514; Nucl. Instr. Methods 205(1983) 47.
\item[{\rm[6]}] F. R. Arutyunian and V. A. Tumanian, Phys. lett. 4, 176(1963);
R. H. Milburn, Phys. Rev. {\bf D47}, 1889(1993).
\item[{\rm[7]}] O. J. P. \'Eboli {\it et al.}, Phys. Rev. {\bf D47},
1889(1993).
\item[{\rm[8]}] J. H. K\"uhn, E. Mirkes, and J. Steegborn, Z. Phys. {\bf C57},
615(1993).
\item[{\rm[9]}] K. Cheung, Phys. Rev. {\bf D47}, 3750(1993); D. Bowser-Chao,
K. Cheung, and S. Thomas, Northewestern preprint NUHEP-TH-93-7(1993).
\item[{\rm[10]}] For a review, see, for example, J. F. Gunion, H. E. Haber, G. Kane
     and S. Dawson, The Higgs Hunters' Guide(Addison-Wesley, Reading, MA,1990).
\item[{\rm[11]}] K. I. Aoki {\it et al.}, Prog. Theor. Phys. Suppl. 73,
1 (1982); M. \"Bohm, W. Hollik, H. Spiesbergerm, Fortschr. Phys. 34, 687 (1986).
\item[{\rm[12]}] V. Telnov, Nucl. Instrum. Methods {\bf A294}, 72(1990);
I. Ginzburg, G. Kotkin, V. Serbo, and V. Telnov, ibid. 2.
\item[{\rm[13]}] Particle Data Group, Phys. Rev. {\bf D50}, No.3(1994).
\item[{\rm[14]}] V. Barger, M. S. Berger and P. Ohmann, Phys. Rev.D47, 
1093(1993).
\item[{\rm[15]}] A. Stange and S. Willenbrock, Phys. Rev. D 48, 2054(1993).
\item[{\rm[16]}] A. Deener, S.Dittmaier and M. Strobel, Preprint, BI-TP ~95/27.
\item[{\rm[17]}] G. Passarino and M. Veltman, Nucl. Phys. B160, 151
(1979);A. Axelrod, Nucl. Phys. B209, 349 (1982); M. Clements $et al.$,
Phys. Rev. D27., 570 (1983). 
\end{description}
\end{itemize}
\vfil
\eject

\begin{center}
{\large Figure Captions}
\end{center}
\begin{description}
\item[\bf Fig.1] ~~Feynman diagrams contributive to $O(\alpha m_t^2/m_W^2)$ Yukawa
	corrections to $\gamma \gamma \rightarrow t \bar t$:
(a),(b) tree level diagrams; (c)-(e) self-energy diagrams;
(f)-(i) vertex diagrams; (j) including neutral Higgs exchange
diagrams; (k) including $\gamma\gamma H^+H^+(\gamma\gamma
 G^+ G^+)$ -coupling diagrams; (l)-(n) box diagrams.
Here we only plot the one-loop diagrams corresponding
to tree-level diagram (a).
The dashed lines represent $H,h,A,H^{\pm},G^0,G^{\pm}$ for diagrams
(c),(d),(e),(f),(h),(l),  $H^{\pm},G^{\pm}$ for diagrams
(g),(i),(k),(m),(n) and $H,h$ for diagrama(j).\\
\item[\bf Fig.2]~~Plot $\Delta\sigma/\sigma_0\,$ versus $M_h$ in the
standard model.\\
\item[\bf Fig.3]~~Plot $\Delta\sigma/\sigma_0\,$ versus $M_h$ for
$M_A=600$ GeV in the two-Higgs-doublet model ($\alpha=\beta=0.25$).\\
\item[\bf Fig.4]~~Plot $\Delta\sigma/\sigma_0\,$ versus $M_A$ for
$M_h=600$ GeV in the two-Higgs-doublet model ($\alpha=\beta=0.25$).\\
\item[\bf Fig.5]~~Plot $\Delta\sigma/\sigma_0\,$ versus $M_A$ in the
minimal SUSY model for $\tan\beta=0.25$.\\
\item[\bf Fig.6]~~Plot $\Delta\sigma/\sigma_0\,$ versus $M_A$ in the
minimal SUSY model for $\tan\beta=1$.\\
\item[\bf Fig.7]~~Plot $\Delta\sigma/\sigma_0\,$ versus $M_A$ in the
minimal SUSY model for $\tan\beta=5$.\\
\end{description}

\end{document}